\documentclass[11pt,pre,onecolumn,nofootinbib
]{revtex4-2}

\usepackage{lipsum}  
\usepackage{amsmath}
\usepackage{amsfonts}
\usepackage{amssymb}
\usepackage{comment}
\usepackage{graphicx}
\usepackage{dcolumn}
\usepackage{bm}
\usepackage{hyperref}

\begin{document}

\preprint{APS/123-QED}

\title{Nonlinear-Cost Random Walk: exact statistics of the distance covered for fixed budget}


\author{Satya N. Majumdar$^1$, Francesco Mori$^2$, Pierpaolo Vivo$^3$}
 \affiliation{$^1$ LPTMS, CNRS, Univ. Paris-Sud, Université Paris-Saclay, 91405 Orsay, France\\ $^2$  Rudolf Peierls Centre for Theoretical Physics, University of Oxford, Oxford, United Kingdom\\ $^3$ Department of Mathematics, King’s College London, London WC2R 2LS, United Kingdom}

\date{\today}

\begin{abstract}

We consider the Nonlinear-Cost Random Walk model in discrete time introduced in [\textit{Phys. Rev. Lett.} {\bf 130}, 237102 (2023)], where a fee is charged for each jump of the walker. The nonlinear cost function is such that slow/short jumps incur a flat fee, while for fast/long jumps the cost is proportional to the distance covered. In this paper we compute analytically the average and variance of the distance covered in $n$ steps when the total budget $C$ is fixed, as well as the statistics of the number of long/short jumps in a trajectory of length $n$, for the exponential jump distribution. These observables exhibit a very rich and non-monotonic scaling behavior as a function of the variable $C/n$, which is traced back to the makeup of a typical trajectory in terms of long/short jumps, and the resulting ``entropy'' thereof. As a byproduct, we compute the asymptotic behavior of ratios of Kummer hypergeometric functions when both the first and last arguments are large. All our analytical results are corroborated by numerical simulations.
\end{abstract}

\maketitle


\section{Introduction}

Stochastic processes have long been indispensable tools for modeling diverse phenomena spanning a multitude of disciplines, encompassing the realms of biology, finance, engineering, and beyond \cite{berg_book,biology,Bouchaud05,moreau09}. These processes, whose simplest incarnation is the classical random walk in discrete time, are useful to model systems undergoing transitions between various states, guided by probabilistic rules. Over the past century, scientists have harnessed the power of stochastic modeling to gain insights into complex dynamical systems, where randomness and uncertainty play a major role.

Intriguingly, many of these systems admit a quite natural description in terms of \emph{costs} (or \emph{rewards}) associated with the transitions between different \emph{states}. These cost functions are often governed by nonlinear functions, as argued below. This coupling of stochastic dynamics with nonlinear cost structures has unveiled unexpected and even paradoxical behaviors, giving rise to fascinating questions and practical applications. 
Consider, for instance, the motion of animals in space in the presence of environmental noise. Several animals employ intermittent strategies \cite{oshanin07}, switching between periods of rapid locomotion and phases of deliberate, slow-paced movement, during which they search locally for food or correct their direction \cite{moreau09,benichou11}. The reduced distance covered during these slower intervals can be characterized as an effective nonlinear cost \cite{peleg16}.

In the realm of wireless communication, devices can operate in different activity/inactivity states (`off,' `idle,' `transmit,' or `receive') with different levels of energy consumption (`costs') when they undergo transitions between these states \cite{wireless,wireless2}. Even in the complex domain of biochemical reactions, which take place sequentially through a series of intermediate (secondary) states (reactions), it is often convenient to associate a total cost to the primary reaction (e.g. the energy/heat produced or consumed overall) being the sum of the intermediate costs generated by the secondary reactions \cite{stochasticreaction}. In the world of finance and risk-management, consider for instance a car driver's insurance premium: in the so-called \emph{bonus-malus} regime, the number of car accidents caused by the driver within a given insurance window will determine a jump in how much money they will be asked to pay (premium) to insure their vehicle in the future. The premium typically follows a highly non-linear pattern, characterized by a steep increase for reckless drivers, and a long recoil period to get back to a more convenient insurance class after an accident \cite{markovreward4}. In the context of software development, time-pressured programmers often face a difficult choice between implementing robust designs and safeguards in their code -- a preferable but more expensive long-term solution -- or adopting an expedient and patchy fix to rush the project forward but incurring in the so-called ``technical debt" \cite{technicaldebt}. In all the examples above, the total cost or reward of different trajectories (i.e. sequences of transitions) between the \emph{same} initial and final states may depend on the precise microscopic arrangement of ``cheap'' vs. ``costly'' jumps due to the non-linear nature of the cost function involved. Additionally, in all these examples it is quite interesting and natural to ask how a global budget constraint (for example, fixing the total energy of a foraging animal or a wireless device, or the total amount of resources to dedicate to a task) will impact the number and type (cheap vs. costly) of the allowed transitions that make up a typical trajectory. In this paper, we provide an analytical answer to this question in the context of a model of diffusion in discrete time and continuous space that we introduced in \cite{letterPRL}.



In the fields of Mathematics and Engineering, the study of stochastic processes entwined with costs has been looked at through the lens of so-called \emph{Markov reward models} \cite{howard_book,markovreward,rew}, where a cost/reward is associated to each jump between the states of a Markov process. 
In \cite{costlevy}, a random walk within a Lévy random environment until to a first-passage event was investigated. Considering a nonlinear cost associated to each step, the joint distribution of displacement and total cost was derived. In Refs.~\cite{debruyne21,CostResetting} stochastic systems with resetting where a cost is associated to each restart were investigated. In particular, in Ref.~\cite{debruyne21}, optimal control theory was applied to identify the optimal resetting strategy to minimize a given cost function. In \cite{CostResetting}, a random walk with constant resetting rate was considered with a space-dependent cost for each resetting event. The statistics of the total cost was computed analytically for a variety of cost functions. In spite of these interesting works, a more thorough exploration of how nonlinear cost functions influence cost fluctuations, both within the typical and large deviation regimes, remains an interesting frontier ripe for further investigation.

In a recent Letter \cite{letterPRL}, we introduced the Nonlinear-Cost Random Walk (NCRW) model in discrete time, and we used the everyday occurrence of taxi rides and associated fares as a prominent motivation for its study. Taxi journeys through bustling cities often entail a mixture of rapid progress and sluggish segments, dictated by factors such as traffic congestion and traffic lights. The fare charged to passengers is algorithmically determined on the fly by a device - the taxi meter - that adheres to a rather universal and straightforward recipe \cite{Eastaway}. Each municipality prescribes a threshold speed, $\eta_c$, derived from statistical analyses of local traffic patterns. If the taxi surpasses $\eta_c$, the meter tallies the fare based on the distance covered, while a slower pace results in time-based fare computation. This seemingly fair approach ensures that drivers are compensated even when they face prolonged periods of slow progress. For example, London's Tariff I rate dictates that the meter should charge 20 pence for every 105.4 meters covered or 22.7 seconds elapsed, whichever is reached first \cite{TFL}. The seemingly innocuous structure of the taxi fare calculation conceals a fascinating phenomenon known as the ``taxi paradox" \cite{Eastaway}. This paradox materializes when two taxis commence their journey together from point A and arrive simultaneously at point B, yet levy substantially different fares due to their unique sequences of slow and fast segments during their trajectories.

\begin{figure}[h]
    \centering
    \fbox{\includegraphics[scale = 0.66]{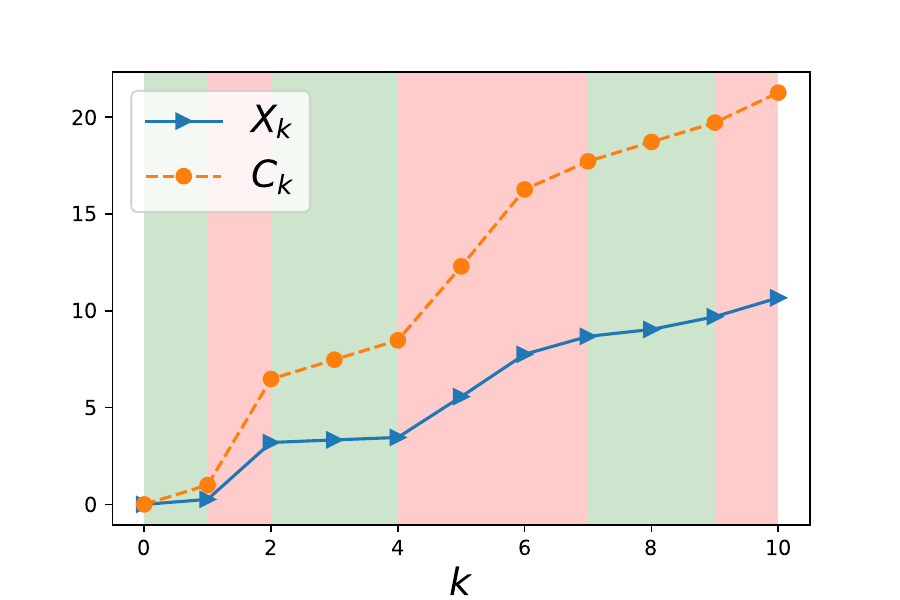}}
    \caption{Typical realization of the system with nonlinearity $h(\eta)=1+b(\eta-\eta_c)\theta(\eta-\eta_c)$ with $b=2$ and $\eta_c=0.7$. The cost $C_k$ up to step $k$ is a nonlinear function of the position $X_k$ of the random walk. Slow (fast) steps with $\eta<\eta_c$ ($\eta>\eta_c$) are highlited with a green (red) background.}
    \label{fig:traj}
\end{figure}

The NCRW model is a Markov process, where a one-dimensional walker's position $X_n$ at discrete time $n$ is a positive random variable evolving according to
\begin{equation}
    X_n = X_{n-1}+\eta_n\label{eq:Xn}
\end{equation}
starting from the origin $X_0=0$. The jumps $\eta_n$ are positive random variables, drawn independently at each time from an exponential pdf $p(\eta)=\exp(-\eta)$. Each jump incurs a positive cost $C_n$, which also evolves via a Markov jump process described by
\begin{equation}
    C_n=C_{n-1}+h(\eta_n)\ ,\label{eq:Cn}
\end{equation}
where $h(\eta)$ is a non-linear cost function, which we take of the form
\begin{equation}
  h(\eta)=1+b(\eta-\eta_c)\theta(\eta-\eta_c)\ ,
  \label{eq:h_eta}
\end{equation}
where $\theta(x)$ is the Heaviside step function. This function $h(\eta)$ mimics the way taxi meters work, since jumps
shorter than the critical size $\eta_c$ in one unit of time (slower
jumps) incur a unit fee, whereas longer (faster) jumps are more costly, with the fee being proportional to the length
(velocity) of the jump. For a typical trajectory of the system, see Fig.~\ref{fig:traj}. Our model has therefore two parameters: $b$, the cost per unit distance covered at high speed, and $\eta_c$, the critical jump length (or speed) separating time-like and space-like charges. The position $X=\sum_{i=1}^n \eta_i$ and the cost $C=\sum_{i=1}^n h(\eta_i)$ after $n$ steps are therefore correlated random variables, whose statistics is of interest.

The kind of non-linearity encoded in the ``cost function'' $h(\eta)$ has many other interesting incarnations in Physics. Consider for instance the force needed to move a block in contact with a surface. One has first to overcome a threshold
force $\eta_c$ due to static friction. Then, applying a force $\eta$ for a fixed time interval $\Delta t$, the velocity of the block is given by $h(\eta)=b(\eta-\eta_c)\theta(\eta-\eta_c)$, where $b$ now depends on the block mass and $\Delta t$ \cite{Vanossi}. Assume now that we repeat this experiment many times, drawing the applied force from, say, an exponential probability density function (pdf), $p(\eta)=\exp(-\eta)$. What would be the average response of the block? It turns out that the mean force per sample $(1/n)\sum_i\eta_i$ over $n$ experiments has the same expression as the final position reached by the taxi in $n$ steps, and the mean velocity of the block per sample precisely corresponds to the average fare for a taxi ride \cite{letterPRL}. Another context where our results could be applied almost directly, is the pinning-depinning transition occurring when an extended object/manifold such as an elastic string or a polymer is
driven by a random force $\eta$ in a spatially inhomogeneous medium \cite{chauve00,duemmer05,Reichhardt16}. Below the depinning threshold $\eta_c$, the manifold is pinned by the disorder and its velocity vanishes, while above the threshold, the velocity-force relation follows a power-law scaling $h(\eta)\propto (\eta-\eta_c)^\beta$, with the depinning exponent $\beta>0$ \cite{chauve00,duemmer05,Reichhardt16}. Examples include DNA chains through nanopores, harmonic elastic strings, and type-II superconductors and colloidal crystals \cite{Blatter94,Pertsinidis08,menais18}.

In our Letter \cite{letterPRL}, we considered two different ensembles of trajectories of a NCRW: in Ensemble (i), we fixed the total distance $X$ covered by the walker as well as the number $n$ of steps, and studied the average $\langle C\rangle_{X,n}$ and variance $\mathrm{Var}(C|X,n)$ of the total cost charged. In Ensemble (ii), instead, we allowed the number of steps $n$ to reach the target destination $X=L$ to fluctuate, and we focused on the hitting cost $C$ -- i.e. the price to pay to reach the destination for the first time -- and its distribution $P(C|L)$ for large $L$. In the former setting [Ensemble (i)], we found a strongly non-monotonic behavior of the variance $\mathrm{Var}(C|X,n)$ as a function of the scaling variable $y=X/(n\eta_c)$, with a maximum attained at the value $y^\star = 1.72724...$ . This non-monotonic behavior was found to reflect a crossover between different phases: (i) a pure phase for small $y=X/(n\eta_c)$, where a typical trajectory is mostly made up of small/slow jumps, (ii) a mixed phase for intermediate values of $y=X/(n\eta_c)$, characterized by a large ``entropy'', i.e. a large number of possible arrangements of the slow and fast jumps to reach the destination $X$, which leads to the strongest fluctuations in the price of the ride, and (iii) again a pure phase for large $y=X/(n\eta_c)$, where a typical trajectory is mostly made up of large/fast jumps, and cost fluctuations from one trajectory and another are suppressed. In Ensemble (ii), where the number of jumps to reach the destination is allowed to fluctuate, we found that the typical fluctuations of hitting cost $C$ are Gaussian but with left and right large deviation tails that can be characterized analytically. The variance of the hitting cost in the typical regime displays a very rich behavior as a function of $b$ -- the cost per unit distance -- and $\eta_c$ -- the critical changeover speed, with a freezing transition in the large deviation regime for $b\eta_c=1$. The resulting giant fluctuations of the hitting cost are once again related to the ``entropic'' makeup of typical trajectories, in terms of the arrangement of short/slow and long/fast jumps to reach the target, and the associated total cost $C$.

In this paper, we consider the NCRW model defined in the equations \eqref{eq:Xn}, \eqref{eq:Cn} and \eqref{eq:h_eta} from a completely different but quite natural perspective, namely by considering global constraints on the total budget available, $C$, and the total number of jumps $n$ to reach a random destination $X$. Stochastic processes with a global constraint have been investigated in the past, and shown to lead to interesting effects. In \cite{condensation1,condensation2}, the distribution of linear statistics of otherwise i.i.d. random variables -- subject to a global constraint on the total ``mass'' -- was shown to undergo an interesting \emph{condensation} transition, where one of the variables acquires a macroscopic fraction of the total mass. In the context of the pinning-depinning transition described above \cite{chauve00,duemmer05,Reichhardt16}, fixing $C$ allows to investigate fluctuations of the total force $X$ at fixed velocity $C$. Moreover, stochastic processes with global nonlinear constraint arise in the context of the discrete nonlinear Schrödinger equation \cite{graden21} as well, where unexpected localization transitions are observed. In this case, the global constraint enforces the conservation of energy in the system. More generally, random models with global constraints are very useful in a wide range of applications where budget constraints are present. This applies, for instance, to macroeconomics \cite{Obstfeld2000,gomez16}, where governments have to allocate resources within a total budget, and supply chain management \cite{beamon98}, where storage space, time, and budget constraints play a crucial role.

We study the average and variance of the distance covered in $n$ steps, as well as the statistics of the number of long/short jumps in a typical trajectory of size $n$, for the exponential jump distribution and assuming that the total budget $C$ for the trajectory is fixed. These quantities display an interesting scaling behavior as a function of the cost per step $C/n$, which we are able to characterize analytically. Furthermore, we find two different regimes, depending on whether $b\eta_c>1$ or $b\eta_c<1$. In the former case, when $b\eta_c>1$, the behavior of the average distance covered as a function of the number of steps $n$ is strongly non-monotonic, which means that actually more steps are needed to cover a shorter distance with the given budget. In the latter case, when $b\eta_c<1$, the curves are instead monotonically decreasing. We give later on a detailed and intuitive explanation for the crossover between the two regimes in terms of the makeup of typical trajectories.  In addition, as a byproduct of our derivations, we determine the asymptotic behavior of ratios of Kummer hypergeometric functions when the first and last argument are both large. Our results are verified by Monte Carlo Markov Chain simulations with a constraint implementing the fixed total budget.

The structure of the paper is as follows. In Section \ref{sec:distance} we consider the statistics of the total distance travelled by the walker in $n$ steps and subject to a budget constraint (total cost equal to $C$): in subsection \ref{sec:PCN} we first compute the pdf $P(C|n)$ of the total cost of a trajectory of $n$ steps (irrespective of the landing spot). This ingredient is needed to compute the constrained average and variance of the final position after $n$ steps, which are tackled in subsections \ref{sec:avX} and \ref{sec:varX}, respectively. In Section \ref{sec:scaling}, we consider the scaling laws obeyed by the constrained average and variance of the final position in the limit $n,C\to\infty$ with their ratio fixed. In Section \ref{sec:longshort}, we consider the statistics of long (fast) vs. short (slow) jumps that make up a typical trajectory, still under the budget constraint. In Section \ref{sec:simul}, we provide the details of the Monte Carlo scheme we employed to simulate trajectories under the fixed-budget constraint. Finally, in Section \ref{sec:conclusions} we offer some concluding remarks. The Appendices are devoted to technical details.


\section{Statistics of distance travelled with fixed budget}\label{sec:distance}

In this section, we focus on $\langle X^k\rangle_{C,n}$, the $k$-th moment of the position reached by the walker after $n$ steps, conditioned on paying a total fare equal to $C$. This quantity represents the total distance travelled by the random walker when both the total budget and the number of steps are fixed.

Consider first the joint pdf of the position $X$ reached after $n$ jumps, and the associated cost $C$, in the case of exponential jump distribution
\begin{equation}
    P(X,C|n)=\int_0^{\infty} d\eta_1\int_0^{\infty}d\eta_2\cdots \int_0^\infty d\eta_n~ e^{-\sum_{i=1}^n\eta_i}\delta\left(X-\sum_i\eta_i\right)\delta\left(C-\sum_i h(\eta_i)\right)\ ,
\end{equation}
where the nonlinear cost function $h(\eta)$ is given in Eq.~\eqref{eq:h_eta}. Taking the double Laplace transform and performing the decoupled $\eta$-integrals, we get
\begin{equation}
    \int_0^\infty\int_0^\infty dX dC ~P(X,C|n)e^{-\lambda X-s C}=G(\lambda,s)=[g(\lambda,s)]^n\ ,\label{doubleLaplace}
\end{equation}
with
\begin{equation}
    g(\lambda,s)=\frac{e^{-s}}{\lambda+1}\left[1-\frac{bs}{\lambda+1+bs}e^{-(\lambda+1)\eta_c}\right]\ .\label{g}
\end{equation}
Taking the $k$-th derivative of \eqref{doubleLaplace} with respect to $\lambda$ and setting $\lambda=0$ we get the Laplace transform (in ``cost'' space) of the $k$-th moment of the final position as
\begin{equation}
    \int_0^\infty dC e^{-s C}\left[\int_0^\infty dX~X^k P(X,C|n)\right]=(-1)^k \frac{\partial^k}{\partial\lambda^k}G(\lambda,s)\Big|_{\lambda=0}\ .\label{momentLaplace}
\end{equation}
For convenience, let us also define
\begin{equation}
    g_k(s)=\frac{\partial^k}{\partial\lambda^k}g(\lambda,s)\Big|_{\lambda=0}\ ,\label{gk}
\end{equation}
with $g_0(s)=g(0,s)$, and $g(\lambda,s)$ given explicitly in \eqref{g}.

Now, from Bayes' theorem, the $k$-th moment of the final position $X$ after $n$ steps -- conditioned on a fixed budget $C$ -- is given by 
\begin{equation}
    \langle X^k\rangle_{C,n}=\frac{\int_0^\infty X^k P(X,C|n)dX}{\int_0^\infty P(X,C|n)dX}\ ,\label{momentdef}
\end{equation}
where the denominator is simply the marginal pdf of the cost alone after $n$ steps
\begin{equation}
 P(C|n)=\int_0^\infty P(X,C|n)dX =\mathcal{L}^{-1}_s \left[G(0,s)\right](C)\label{marginalcost}
\end{equation}
from \eqref{doubleLaplace}. The numerator of \eqref{momentdef} follows by taking the inverse Laplace transform w.r.t. $s$ of Eq. \eqref{momentLaplace}, which reads using Bromwich formula
\begin{equation}
    \int_0^\infty X^k P(X,C|n)dX =(-1)^k
\int_\Gamma \frac{ds}{2\pi\mathrm{i}}e^{sC}g_k(s)\ ,\end{equation}
with $g_k(s)$ defined in \eqref{gk}, and $\Gamma$ a vertical line in the complex plane to the right of all the singularities of the integrand. So, explicitly
\begin{equation}
    \langle X^k\rangle_{C,n}=\frac{(-1)^k \int_\Gamma \frac{ds}{2\pi\mathrm{i}}e^{sC}\frac{\partial^k}{\partial\lambda^k}G(\lambda,s)\Big|_{\lambda=0}}{\int_\Gamma \frac{ds}{2\pi\mathrm{i}}e^{sC}G(0,s)}\ ,\label{generalformulamoment}
\end{equation}
with $G(\lambda,s)=[g(\lambda,s)]^n$ and $g(\lambda,s)$ given in \eqref{g}. Let us now specialize \eqref{generalformulamoment} to the first two moments $k=1$ and $k=2$, after computing the marginal distribution of the total cost alone in the next sub-section.

\subsection{Calculation of $P(C|n)$}\label{sec:PCN}

\begin{figure}
    \centering
    \fbox{\includegraphics[scale = 0.56]{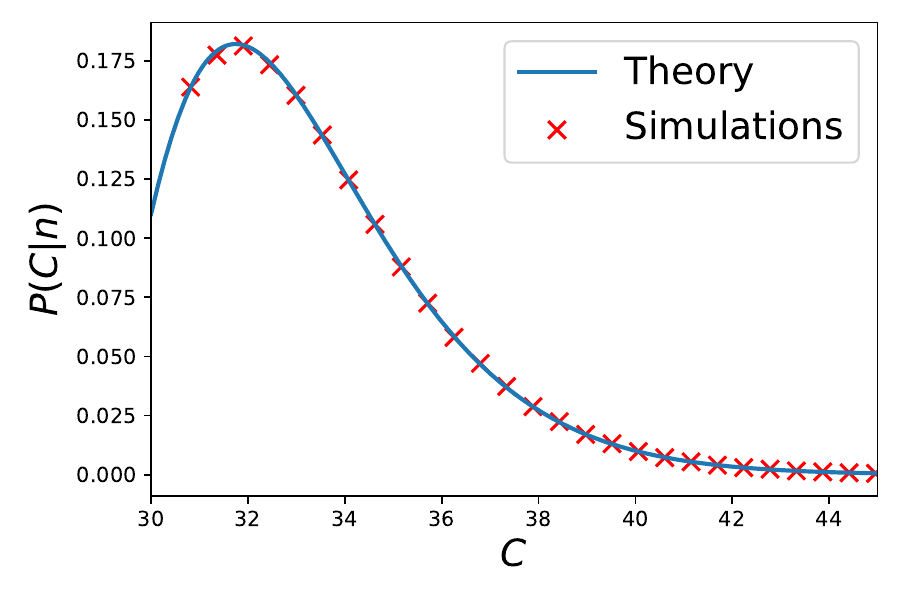}}
    \caption{Probability density function $P(C|n)$ of the total cost after $n$ steps for $C>n$. The blue line corresponds to the theoretical result in Eq.~\eqref{PCNeq} with $n=30$, $b=1$, and $\eta_c=2.2$. The red crosses are obtained from numerical simulations.}
    \label{fig:PCN}
\end{figure}

From \eqref{marginalcost}, we need to compute the inverse Laplace transform (over the variable $s$) of $G(0,s)=[g(0,s)]^n=e^{-ns}[1-e^{-\eta_c}+e^{-\eta_c}/(1+bs)]^n$.
We can rewrite
\begin{equation}
    G(0,s)=(1-e^{-\eta_c})^n e^{-ns}\left[1+\frac{A}{1+bs}\right]^n=(1-e^{-\eta_c})^n e^{-ns}\left[1+\sum_{k=1}^n {n\choose k}\frac{A^k}{(1+bs)^k}\right]\ ,
\end{equation}
where 
\begin{equation}
    A=(e^{\eta_c}-1)^{-1}\,. \label{eq:A}
\end{equation}
The function $G(0,s)$ can then be easily Laplace-inverted term by term using
\begin{align}
    \mathcal{L}^{-1}_s [e^{-ns}](C) &=\delta(C-n)\\
    \mathcal{L}^{-1}_s \left[\frac{e^{-ns}}{(1+bs)^k}\right](C) &=\frac{e^{-\frac{C-n}{b}} \theta (C-n) \left(\frac{C-n}{b}\right)^{k-1}}{b \Gamma (k)}\ ,
\end{align}
with $\theta(x)$ the Heaviside step function. Using next the identity
\begin{equation}
    \sum_{k=1}^n{n\choose k}\frac{A^k \left(\frac{C-n}{b}\right)^{k-1}}{\Gamma(k)}=A n \, _1F_1\left(1-n;2;\frac{-A (C-n)}{b}\right)
\end{equation}
in terms of a Kummer hypergeometric function,
\begin{equation}
    _1F_1(a;b; z)= 1+ \frac{a}{b}z + \frac{a(a+1)}{b(b+1)} \frac{z^2}{2!}+\cdots\ ,
\end{equation}
we eventually get
\begin{equation}
    P(C|n)=(1-e^{-\eta_c})^n \delta(C-n)+\frac{n}{b}e^{-\frac{C-n}{b}}\theta(C-n)e^{-\eta_c}(1-e^{-\eta_c})^{n-1}\, _1F_1\left(1-n;2;\frac{-A (C-n)}{b}\right)\ .\label{PCNeq}
\end{equation}
The first delta term is easy to understand: a total cost exactly equal to $n$ can be realized by a sequence of $n$ time-like steps, each of which is charged one unit of cost. A trajectory of $n$ time-like steps occurs with probability $(1-e^{-\eta_c})^n$. A plot of the continuous part of $P(C|n)$ for $C>n$ is included in Fig. \ref{fig:PCN}.
Normalization of $P(C|n)$ over $C\in [n,+\infty)$ can be easily checked using the integral formula 
\begin{equation}
    \int_0^\infty dx~e^{-x}\, _1F_1\left(1-n;2;-A  x\right)=\frac{(1+A)^n-1}{An}\ ,
\end{equation}
after making the change of variables $(C-n)/b = x$.

\subsection{Average of the position reached after $n$ steps on a fixed budget (finite $n$)}\label{sec:avX}

\begin{figure}
    \centering
    \fbox{\includegraphics[scale = 0.5]{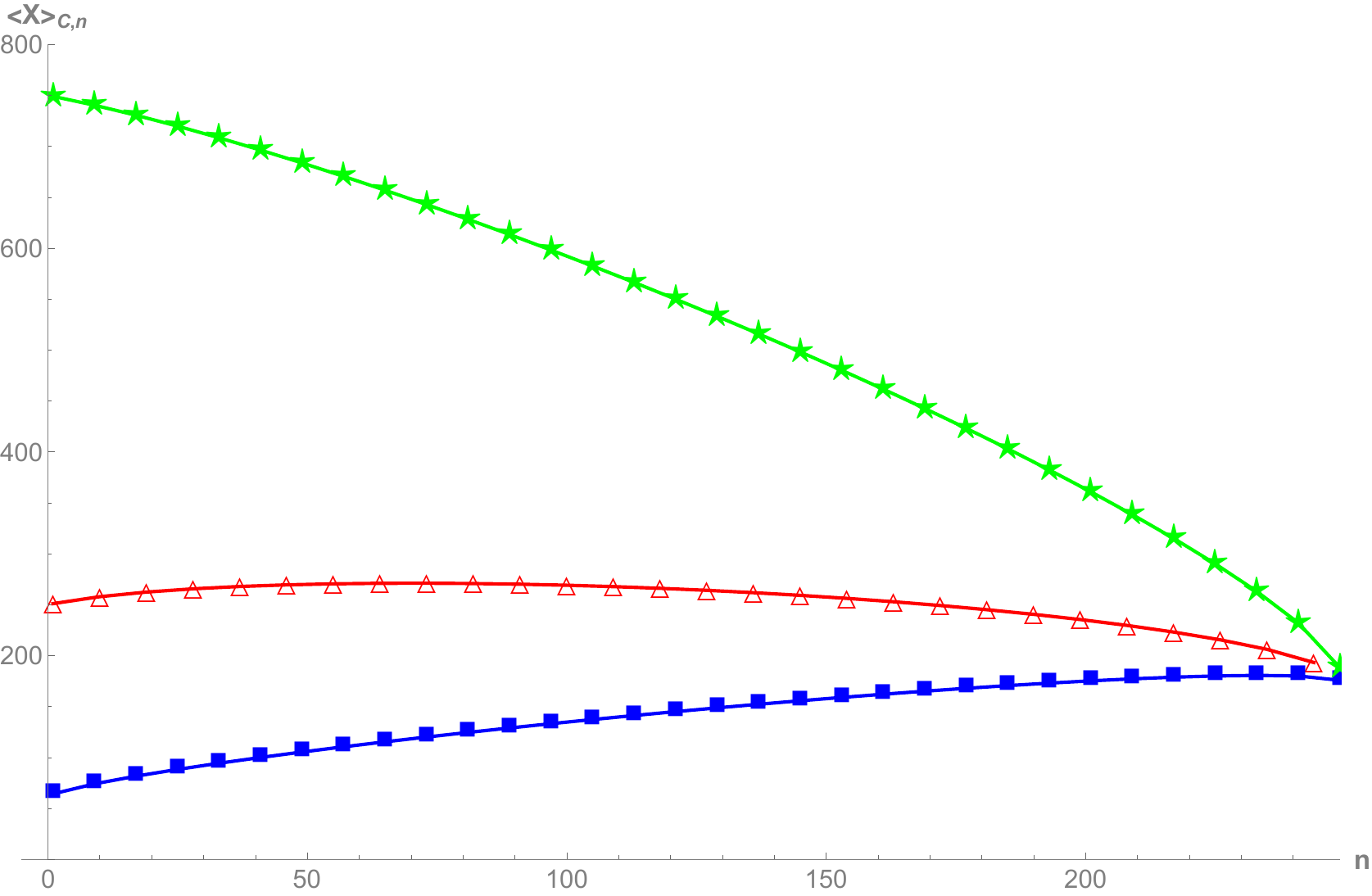}}
    \caption{Average distance $\langle X\rangle_{C,n}$ covered as a function of the number of jumps $n$ for a fixed total budget $C=250$. The critical changeover speed is set at $\eta_c=2$. The blue squares are for $b=4$, the red triangles are for $b=1$, and the green stars are for $b=1/3$. We observe that the first two curves are non-monotonic as a function of $n$, whereas the green curve is monotonically decreasing as $b\eta_c<1$. In the region where $\langle X\rangle_{C,n}$ is decreasing with $n$, more steps lead to a shorter total distance covered. This effect is a consequence of the non-linearity of the cost function, as explained in the text. }
    \label{fig:Avdifferentb}
\end{figure}

Having computed the denominator of \eqref{momentdef}, we now turn to the calculation of the numerator for $k=1$. For convenience, let us denote by
\begin{equation}
    T_1(C)=\langle X\rangle = \int_0^\infty X P(X,C|n)dX\label{eq:T1Cmain}
\end{equation}
the unconstrained average of the final position of the walker after $n$ steps. The calculation from the Laplace transform in Eq. \eqref{momentLaplace} for $k=1$ is reported in Appendix \ref{app:T1C} and yields eventually

\begin{align}
 \nonumber   T_1(C) &=\frac{1}{b}e^{-C/b}\hat T_1 (C/b)=
    \mathcal{A}(n,b,\eta_c)\delta(C-n)+\theta(C-n)e^{-\frac{C-n}{b}}\\
    & \times\left\{\mathcal{B}(n,b,\eta_c) ~ _1F_1\left(2-n;2;-\frac{A(C-n)}{b}\right)+\mathcal{C}(n,b,\eta_c)~ _1F_1\left(1-n;1;-\frac{A(C-n)}{b}\right) \right.\nonumber \\
     &+\left.\mathcal{D}(n,b,\eta_c)(C-n)~ _1F_1\left(1-n;2;-\frac{A(C-n)}{b}\right)\right\}\ ,\label{eqT1C}
\end{align}
where $A$ is given in Eq.~\eqref{eq:A} and
\begin{align}
   \mathcal{A}(n,b,\eta_c) &=n(1-(\eta_c+1)e^{-\eta_c})(1-e^{-\eta_c})^{n-1}\,,\\
   \mathcal{B}(n,b,\eta_c) &=\frac{n(n-1)}{b}(1-e^{-\eta_c})^{n-2}e^{-\eta_c}[1-e^{-\eta_c}(1+\eta_c)]\,,\\
   \mathcal{C}(n,b,\eta_c) &=\frac{n}{b}(1-e^{-\eta_c})^{n-1}\eta_c e^{-\eta_c}\,,\\
   \mathcal{D}(n,b,\eta_c) &=\frac{n}{b^2}e^{-\eta_c}(1-e^{-\eta_c})^{n-1}\ .
\end{align}
Taking the ratio between $T_1(C)$ in \eqref{eqT1C} and the pdf in \eqref{PCNeq} gives the average final position after $n$ steps constrained on a fixed cost $C$ (see Eq. \eqref{momentdef} for $k=1$)
\begin{align}
  \nonumber  \langle X\rangle_{C,n} &=\mathcal{K}_1(n,b,\eta_c)\delta(C-n)+\theta(C-n)\left\{\mathcal{K}_2(n,b,\eta_c)R\left(2,n,2,\frac{A(C/n-1)}{b}\right)\right.\\
    &\left. + \mathcal{K}_3(n,b,\eta_c)R\left(1,n,1,\frac{A(C/n-1)}{b}\right)+\mathcal{K}_4(n,b,\eta_c)R\left(1,n,2,\frac{A(C/n-1)}{b}\right)\right\}\ ,\label{eq:avXconstrainedfinal}
\end{align}
where $A$ is given in Eq.~\eqref{eq:A} and
\begin{equation}
    R(k,n,m,u)=\frac{ _1 F_1(k-n, m, -n u)}{_1 F_1(1-n,2,-nu)}\label{eq:ratioKummer}
\end{equation}
and the constants $\mathcal{K}_j$ can be easily reconstructed. This constrained average is plotted as a function of $n$ in Fig. \ref{fig:Avdifferentb} for $C=250$ and three different values of $b$, the cost per unit distance in the ``high speed'' regime. Quite counter-intuitively, there are parameter choices for which the behavior of the constrained average is strongly non-monotonic as a function of the number of steps $n$: this means that the walker may actually perform \emph{more} jumps to cover a \emph{shorter} distance (on average). The reason is that -- at fixed budget $C$, and with ``too many'' jumps to perform -- the walker is forced to slow its pace down, and burn money on short (time-like) jumps, otherwise the budget would be all spent on ``too few'' (but large) excursions.

\begin{figure}
    \centering
    \fbox{\includegraphics[scale = 0.5]{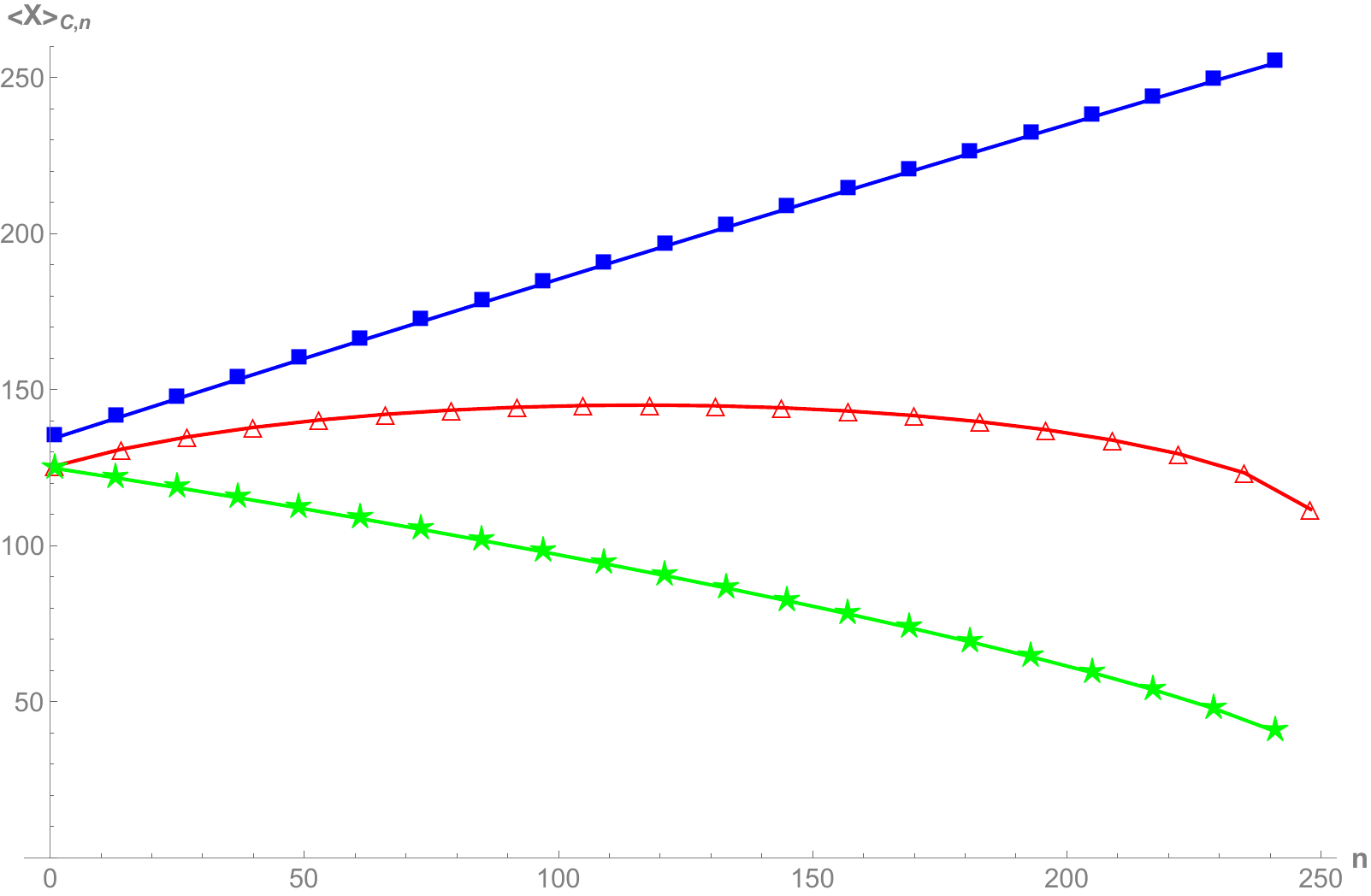}}
    \caption{Average distance $\langle X\rangle_{C,n}$ covered as a function of the number of jumps $n$ for a fixed total budget $C=250$. The cost per unit distance in the ``high speed'' regime is set at $b=2$. The blue squares are for $\eta_c=10$, the red triangles are for $\eta_c=1$, the green stars are for $\eta_c=1/4$. The curve with $\eta_c=1$ displays the non-monotonic effect described in Fig.~\ref{fig:Avdifferentb}. In the case $\eta_c=10$, the probability of a ``high speed'' step ($\eta>\eta_c)$ is extremely low and hence most steps have a unit cost, leading to the observed linear behavior. The green curve is for $b\eta_c<1$ and is therefore monotonically decreasing, as explained in the main text.}
    \label{fig:Avdifferenteta}
\end{figure}

In Fig.~\ref{fig:Avdifferenteta} we plot $\langle X\rangle_{C,n}$ for different values of $\eta_c$ and $b=2$. 
It is easy to understand the initial growth of the red and blue curves as a function of $n$
from the following very simple argument, which also shows that there should be a transition at $b\eta_c=1$, with the curve for $\langle X\rangle_{C,n}$ growing with $n$ if $b\eta_c>1$ and decreasing if $b\eta_c<1$. 

Consider increasing $n$, for fixed but large budget $C$. We recall that the cost function is given by
\begin{equation}
C= \sum_{k=1}^n h(\eta_k)\ ,
\end{equation}
where 
\begin{equation}
h(\eta)= 1+ b (\eta-\eta_c) \theta(\eta-\eta_c)\ .
\end{equation}

When $n=1$, we have only one space-like step, hence, $C= h(\eta_1)= 1+ b(\eta_1-\eta_c)$ (assuming $\eta_1>\eta_c$). But the final destination in this case is $X= \eta_1$, hence, $X= C/b + (b \eta_c-1)/b$. Now, suppose we have $n=2$, and imagine both jumps are space-like. Then
$C= 1+ b(\eta_1-\eta_c) + 1+ b(\eta_2-\eta_c)$, implying $X= \eta_1+\eta_2=  C/b+ 2 (b \eta_c-1)/b$. In general, if we have $n$ space-like steps, then given a large budget $C$
\begin{equation}
    X= \eta_1+\eta_2+\ldots +\eta_n= C/b + n (b\eta_c-1)/b\ .\label{eq:X_sum_etas}
\end{equation}
Thus, we see that if $b\eta_c>1$, $X$ will increase with $n$ initially, as long as the jumps
are space-like. Beyond the maximum (when $b\eta_c>1$), time-like steps start to kick in
and clearly $X$ then have to decrease. This argument explains the non-monotonicity of $\langle X\rangle_{C,n}$ for $b\eta_c>1$. For $b\eta_c<1$, when the slope in Eq.~\eqref{eq:X_sum_etas} becomes negative, the curve decreases monotonically instead, since whether the jumps are space or time-like, $X$ always decreases with increasing $n$ for fixed large $C$.

\subsection{Variance of the position reached after $n$ steps on a fixed budget (finite $n$)}\label{sec:varX}

We now turn to the calculation of the numerator of \eqref{momentdef} for $k=2$. For convenience, let us denote by
\begin{equation}
    T_2(C)=\langle X^2\rangle = \int_0^\infty X^2 P(X,C|n)dX\label{eq:T2Cmain}
\end{equation}
the unconstrained second moment of the final position of the walker after $n$ steps. The calculation is reported in Appendix \eqref{app:T2C}, with the final (long) expression given in Eqs. \eqref{eq:T2C} and \eqref{eq:t2hat}.

The second moment of $X$ -- constrained on a fixed budget $C$ -- can then be computed by taking the ratio of $T_2(C)$ in Eq.~\eqref{eq:T2C} to the pdf $P(C|n)$ in Eq.~\eqref{PCNeq}. We obtain
\begin{align}
    \langle X^2\rangle_{C,n}=\frac{e^{-C/b}}{1-e^{-\eta_c}}M(n,b,\eta_c)\delta(C-n) +\frac{\theta(C-n)e^{-C/b} \hat T_2(C/b)}{ne^{-\frac{C-n}{b}}e^{-\eta_c}(1-e^{-\eta_c})^{n-1}\, _1F_1\left(1-n;2;\frac{-A (C-n)}{b}\right)}\ ,
    \label{eq:x2_final}
\end{align}
where $A$ and $\hat T_2(\hat C)$ are respectively given in Eqs.~\eqref{eq:A} and \eqref{eq:t2hat}, and we defined
\begin{align}
M(n,b,\eta_c) &=  n(n-1)e^{n/b}(1-e^{-\eta_c})^{n-2}(e^{-\eta_c}(1+\eta_c)-1)^2\ .\label{eq:Mnb}
\end{align}
Here, we are using the convention $\delta(C-n)\theta(C-n)=0$. Finally, we obtain 
\begin{equation}
    \operatorname{Var}(X)_{C,n}=\langle X^2\rangle_{C,n}-\langle X\rangle_{C,n}^2\,, \label{eq:Var_final}
\end{equation}
where $\langle X\rangle_{C,n}$ and $\langle X^2\rangle_{C,n}$ are given in Eqs.~\eqref{eq:avXconstrainedfinal} and \eqref{eq:x2_final}. The variance of $X$ is shown in Fig.~\eqref{fig:CondVar} as a function of $n/C$ for $b=1$, $\eta_c=2.1$ and different values of $C$. Our analytical result is in good agreement with numerical simulations (see the inset in Fig.~\eqref{fig:CondVar}). The variance reaches a maximum value at some intermediate value of $y=n/C$ and then decreases for increasing $n/C$. The origin of this non-monotonic behavior is similar to that described in Ref.~\cite{letterPRL} and is the result of an entropic effect. When $n\ll C$, the cost is concentrated in a few ``spacelike'' fast steps and hence the variance is low. In the opposite limit $n\approx C$, most steps are ``timelike'' ($\eta<\eta_c$). At intermediate values of $n$, a mixture of the two type of steps is present, leading to a maximum in the variance. 

\section{Scaling laws}\label{sec:scaling}

In this section, we show (analytically and numerically) that the finite $n,C$ results in the previous sections admit nice scaling laws for large $n$ and large $C$, keeping their ratio $n/C$ fixed. There are two ways to perform this asymptotics: in the next subsection, we work directly in Laplace space, and perform a saddle-point analysis of the ratio of Bromwich integrals in \eqref{generalformulamoment} for $k=1$. Otherwise, in Appendix \ref{app:ratiohyp}, we directly compute the asymptotics of Eq. \eqref{eq:avXconstrainedfinal}, which in turn involves computing the asymptotics of the ratio of Kummer functions in \eqref{eq:ratioKummer} when two of the arguments are large. This asymptotic calculation is a nice byproduct of our work.

\subsection{Average of the position reached after $n$ steps on a fixed budget (scaling formula for large $n$)}

We start from Eq. \eqref{generalformulamoment}
\begin{equation}
    \langle X^k\rangle_{C,n}=\frac{(-1)^k \int_\Gamma \frac{ds}{2\pi\mathrm{i}}e^{sC}\frac{\partial^k}{\partial\lambda^k}G(\lambda,s)\Big|_{\lambda=0}}{\int_\Gamma \frac{ds}{2\pi\mathrm{i}}e^{sC}G(0,s)}\ ,\label{generalformulamomentV2}
\end{equation}
where $G(\lambda,s)=[g(\lambda,s)]^n$, with $g(\lambda,s)$ given in \eqref{g}. For $k=1$, we therefore need to compute
\begin{equation}
    \frac{\partial}{\partial\lambda}[g(\lambda,s)]^n=n [g(\lambda,s)]^{n-1}\frac{\partial}{\partial\lambda}g(\lambda,s)\ ,\label{firstder}
\end{equation}
and setting $\lambda=0$
\begin{equation}
    \frac{\partial}{\partial\lambda}[g(\lambda,s)]^n\Big|_{\lambda=0}=n [g_0(s)]^{n-1}g_1(s)=n [g_0(s)]^{n}\frac{g_1(s)}{g_0(s)}\ ,
\end{equation}
with $g_k(s)$ defined in \eqref{gk}. Hence from \eqref{generalformulamoment} we have for $k=1$
\begin{equation}
    \langle X\rangle_{C,n}=n\frac{\int_\Gamma \frac{ds}{2\pi\mathrm{i}}e^{sC}[g_0(s)]^{n}\frac{-g_1(s)}{g_0(s)}}{\int_\Gamma \frac{ds}{2\pi\mathrm{i}}e^{sC}[g_0(s)]^n}\ .\label{k1}
\end{equation}
Both the denominator and the numerator can be evaluated by the saddle-point method for large $n$. Consider first the denominator and rewrite as
\begin{equation}
   \int_\Gamma \frac{ds}{2\pi\mathrm{i}}e^{sC}[g_0(s)]^n=\int_\Gamma \frac{ds}{2\pi\mathrm{i}}e^{n\left[s \frac{C}{n}+\ln g_0(s)\right]}\ . \label{SPden}
\end{equation}
Setting $C/n=z$ (fixed for large $n$), the action in the exponent reads
\begin{equation}
    A(s,z)=sz+\ln g_0(s)\ .
\end{equation}
The stationary point of the action is determined by
\begin{equation}
    \frac{\partial A(s,z)}{\partial s}=0\Rightarrow z+\frac{g_0'(s)}{g_0(s)}=0\ ,\label{SP}
\end{equation}
whose solution implicitly provides the critical value $s=s^\star(z)$.

Consequently, from \eqref{SPden}\footnote{The symbol $\approx$ denotes asymptotic equality on logarithmic scales.}
\begin{equation}
    \int_\Gamma \frac{ds}{2\pi\mathrm{i}}e^{sC}[g_0(s)]^n\approx \exp\left[n\Phi(z)\right]
\end{equation}
for $n\to\infty$, $C\to\infty$ such that $z=C/n$ is fixed, with
\begin{equation}
    \Phi(z)=\max_s \left[sz+\ln g_0(s)\right]=s^\star(z)z+\ln g_0(s^\star(z))\ .
\end{equation}

Looking back at \eqref{k1}, since $g_1(s)/g_0(s)$ is independent of $n$, to leading order for large $n$ the numerator will be dominated by the behavior in the vicinity of the very same saddle point as the denominator, namely $s^\star(z)$. Therefore, the leading exponential terms will cancel out, and what remains is
\begin{equation}
    \langle X\rangle_{C=nz,n}\sim -n\left[\frac{g_1(s^\star(z))}{g_0(s^\star(z))}\right]\label{initialscaling}
\end{equation}
for large $n$.

Now, defining $r(s)=1+bs$, we have that $g_0(s)$ is given by \eqref{g} as
\begin{equation}
    g_0(s)=e^{-s}\left[1-e^{-\eta_c}+\frac{e^{-\eta_c}}{r(s)}\right]\ ,
\end{equation}
while
\begin{equation}
    g_1(s)=e^{-s}\left[bs e^{-\eta_c}\left(\frac{1}{r(s)^2}+\frac{\eta_c+1}{r(s)}\right)-1\right]\ .
\end{equation}
For the saddle point condition \eqref{SP} we get
\begin{equation}
    -z=\frac{g_0'(s^\star)}{g_0(s^\star)}=-1-\frac{\frac{e^{-\eta_c}b}{r(s^\star)^2}}{1-e^{-\eta_c}+\frac{e^{-\eta_c}}{r(s^\star)}}\ ,\label{SPexplicit}
\end{equation}
which gives the following quadratic equation for $r^\star=r(s^\star)$
\begin{equation}
    (e^{\eta_c}-1)r^{\star 2}+r^\star-\frac{b}{z-1}=0\ .
\end{equation}
Its positive root\footnote{Since $s>0$, $r>0$ as well.} $r^\star$ reads
\begin{equation}    r^\star(z)=\frac{-1+\sqrt{1+4(e^{\eta_c}-1)\frac{b}{z-1}}}{2(e^{\eta_c}-1)}\ .
\end{equation}

\begin{figure}
    \centering
    \fbox{\includegraphics[scale = 0.5]{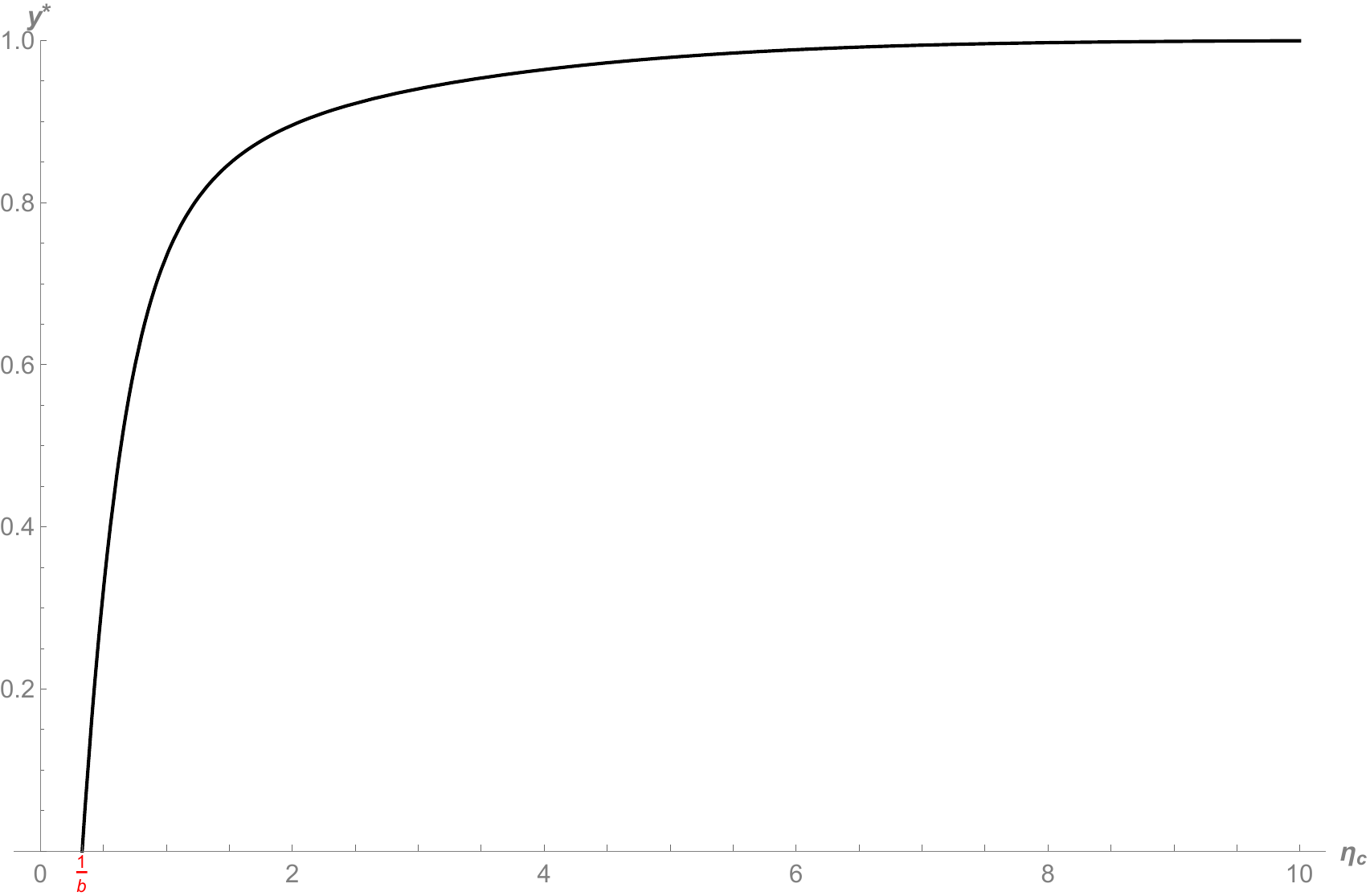}}
    \caption{Behavior of $y^\star=\mathrm{argmax}f(y)$ in Eq.~\eqref{eq:ystar} as a function of $\eta_c$ for $b=3$. For $\eta_c<1/b$, the maximum is reached at the lower edge $y=0$. }
    \label{fig:maximum}
\end{figure}

\begin{figure}
    \centering
    \fbox{\includegraphics[scale = 0.4]{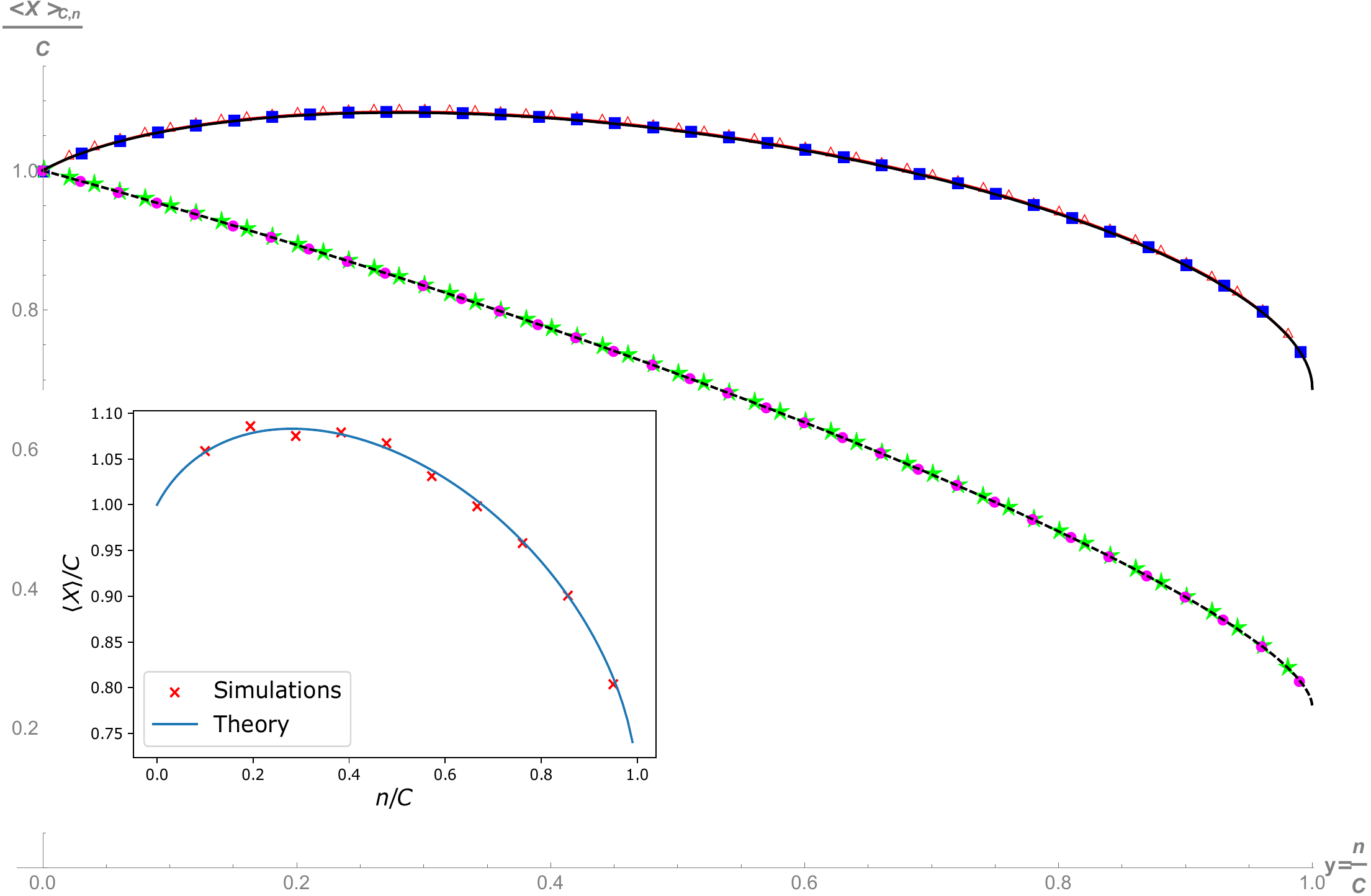}}
    \caption{Scaling behavior of the average final position $\langle X\rangle_{C,n}$ after $n$ steps conditioned on the total cost of the ride. On the vertical axis, we have the quantity $\langle X\rangle_{C,n}/C$ from Eq. \eqref{eq:avXconstrainedfinal}. On the horizontal axis, we put $y=n/C$. We use $b=1$, $\eta_c=2$, and $C=250$ (red triangles) or $C=100$ (blue squares). We use instead $b=1$, $\eta_c=1/2$, and $C=250$ (green stars) or $C=100$ (magenta circles). We can see that the finite $n,C$ curves nicely collapse onto the scaling function $f(y)$ given in Eq. \eqref{eq:scaling_f}. The scaling curves show a non-monotonic behavior for $b\eta_c>1$, and a monotonic (decreasing) behavior for $b\eta_c<1$, as explained in the main text. Inset: Average displacement $\langle X\rangle_{C,n}/C$ as a function of $n/C$ for $\eta_c=2$, $b=1$, and $C=100$. The red crosses correspond to the results of numerical simulations (see Sec.~\ref{sec:simul} for the details), while the continuous blue line corresponds to the scaling form in Eq.~\eqref{eq:scaling_f}. }
    \label{fig:CondAv}
\end{figure}

Summarizing, from \eqref{initialscaling}, it follows that the average position reached after $n$ steps and constrained on a fixed total budget $C$ has the following scaling form for large $n$ (and noting that $y=1/z$)
\begin{equation}
     \langle X\rangle_{C,n}\sim C f\left(y=\frac{n}{C}\right)\ , \label{finalscaling}
\end{equation}
where the scaling function 
\begin{equation}
    f(y)=\frac{1-y}{b}\left[\tilde r(y)^2 (e^{\eta_c}-\eta_c -1)+\eta_c  \tilde r(y)+1\right]\ ,\label{eq:scaling_f}
    \end{equation}
with 
\begin{equation}
  \tilde r(y)= \frac{A}{2} \left[\sqrt{\frac{4 b y}{A(1-y)}+1}-1\right]
\end{equation}
and $0\leq y\leq 1$ (as the budget for a ride of $n$ steps can never be smaller than $n$). We recall that $A$ is defined in Eq.~\eqref{eq:A}. The asymptotic behaviors are as follows
\begin{equation}
    f(y)\sim\begin{cases}
        1/b &\qquad\text{for }y\to 0^+\\
        A(-\eta_c +e^{\eta_c }-1) &\qquad\text{for }y\to 1^-\ .
    \end{cases}
\end{equation}
The scaling function has an interesting non-monotonic behavior (unless $\eta_c<1/b$) as a function of $y$, with a maximum at the value 
\begin{equation}
    y^*=\frac{2b(e^{\eta_c}-1)-1+\frac{b(-1+2b(e^{\eta_c}-1)(e^{\eta_c}-\eta_c-1)+e^{\eta_c}(3-2e^{\eta_c}+\eta_c))}{\sqrt{e^{\eta_c}(e^{\eta_c}-\eta_c-1)+b^2(e^{\eta_c}-\eta_c-1)^2+b(-2-\eta_c+e^{\eta_c}(4-2e^{\eta_c}+\eta_c+\eta_c^2))}}}{4b(e^{\eta_c}-1)}\,.\label{eq:ystar}
\end{equation}
The value $y^\star$ is plotted as a function of $\eta_c$ for $b=3$ in Fig. \ref{fig:maximum}, while in Fig. \ref{fig:CondAv} we plot two instances of the scaling curve $f(y)$ (for $b\eta_c>1$ and $b\eta_c<1$), compared with the finite $n,C$ behavior from Eq. \eqref{eq:avXconstrainedfinal}. We clearly observe once again a different behavior of the curves if $b\eta_c>1$ or $b\eta_c<1$, as argued at the end of Section \ref{sec:avX}. This is reflected in the fact that for $b\eta_c<1$, the maximum $y^\star$ of the curve is reached at the lower edge $y=0$. In the inset of Fig.~\eqref{fig:CondAv} we compare our theoretical result in Eq.~\eqref{finalscaling} with numerical simulations, finding excellent agreement.

\begin{figure}
    \centering
    \fbox{\includegraphics[scale = 0.5]{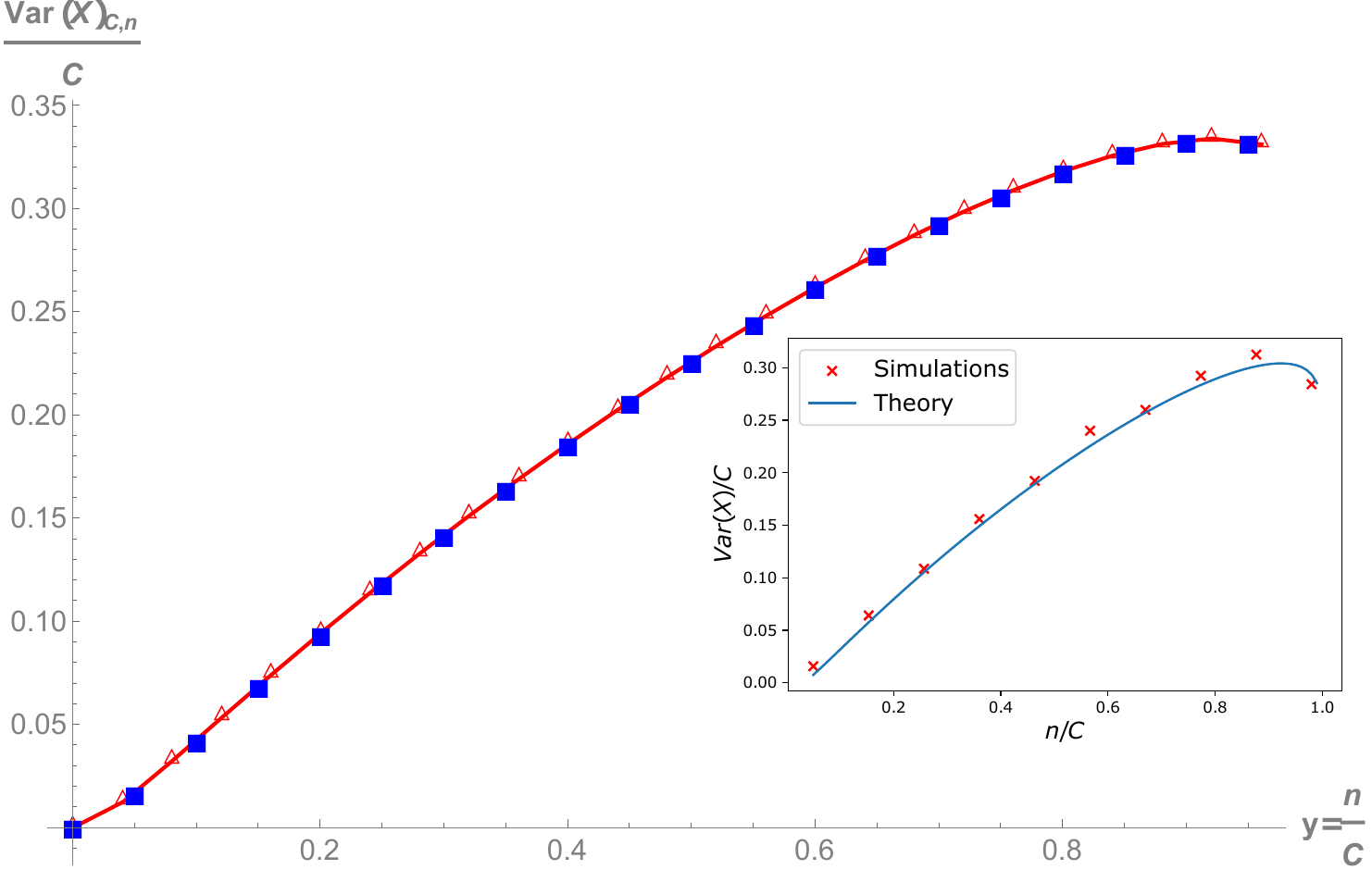}}
    \caption{Scaling behavior of the variance of the final position $\mathrm{Var}(X)_{C,n}$ after $n$ steps conditioned on the total cost of the ride. On the vertical axis, we have the quantity $\mathrm{Var}(X)_{C,n}/C$ from Eq.~\eqref{eq:Var_final}. On the horizontal axis, we put $y=n/C$. We use $b=1$, $\eta_c=2.1$, and $C=250$ (red triangles, joined with a solid line) or $C=100$ (blue squares). We can see that the finite $n,C$ curves nicely collapse onto a scaling function, which would be too cumbersome to determine explicitly. Inset: Scaled variance $\operatorname{Var}(X)_{C,n}/C$ as a function of $n/C$. The red crosses are the result of numerical simulations with $C=30$, $\eta_c=2$, and $b=1$. The blue line corresponds to the exact result in Eq.~\eqref{eq:Var_final}.}
    \label{fig:CondVar}
\end{figure}

\subsection{Variance of the position reached after $n$ steps on a fixed budget}

Similarly, we could compute the second moment (and thus the variance) of the position reached after $n$ steps on a fixed budget $C$ by setting $k=2$ in \eqref{generalformulamoment}. First, taking a further derivative w.r.t. $\lambda$ of \eqref{firstder}, we get
\begin{equation}
    \frac{\partial^2}{\partial\lambda^2}[g(\lambda,s)]^n=n(n-1)[g(\lambda,s)]^{n-2}\left(\frac{\partial g}{\partial\lambda}\right)^2+n[g(\lambda,s)]^{n-1}\frac{\partial^2 g}{\partial \lambda^2}\ .
\end{equation}
Setting $\lambda=0$, and using \eqref{gk}, we get
\begin{equation}
    \frac{\partial^2}{\partial\lambda^2}[g(\lambda,s)]^n\Big|_{\lambda=0}=n(n-1)[g_0(s)]^n \left[\frac{g_1(s)}{g_0(s)}\right]^2+n [g_0(s)]^n \frac{g_2(s)}{g_0(s)}\ .
\end{equation}
Hence from \eqref{generalformulamoment}
\begin{equation}
    \langle X^2\rangle_{C,n}=\frac{ \int_\Gamma \frac{ds}{2\pi\mathrm{i}}e^{sC}[g_0(s)]^n \left[n(n-1) \left[\frac{g_1(s)}{g_0(s)}\right]^2+n  \frac{g_2(s)}{g_0(s)}\right]}{\int_\Gamma \frac{ds}{2\pi\mathrm{i}}e^{sC}[g_0(s)]^n}\ . \label{eq:X2avg}
\end{equation}

While we could again use a saddle-point evaluation of both numerator and denominator for large $n$, it turns out that it would not be enough to confine the analysis to the leading $\sim n^2$ term to extract the leading term of the constrained variance. Extracting the sub-leading term $\sim n$ requires a careful and very laborious calculation, which would not be rewarded by a particularly illuminating final result. For these reasons, we decided to show the scaling behavior of the constrained variance only numerically in Fig. \ref{fig:CondVar}.

\section{Statistics of the number of short/long jumps}\label{sec:longshort}

In this section, we consider the statistics of the random variable
\begin{equation}
    n_t = \sum_{i=1}^n \theta(\eta_c-\eta_i)
\end{equation}
where $\theta(x)$ is the Heaviside step function. The random variable $n_t$ counts the number of time-like jumps in a trajectory of length $n$. The joint pdf of $n_t$, the final position $X$ and the total cost (budget) $C$ for a trajectory of length $n$ and exponential jump distribution is given by
\begin{align}
    P(n_t,X,C|n)&=\int_{(0,\infty)^n}d\eta_1\cdots d\eta_n e^{-\sum_{i=1}^n\eta_i}\delta\left(X-\sum_i\eta_i\right)\delta\left(C-\sum_i h(\eta_i)\right)\nonumber \\
    &\times \delta\left(n_t - \sum_{i=1}^n \theta(\eta_c-\eta_i)\right)\ .
\end{align}
Taking the triple Laplace transform
\begin{equation}
    \int dn_t dX dC~P(n_t,X,C|n)e^{-\lambda X-s C- \xi n_t}=[\chi(\lambda,s,\xi)]^n\ ,\label{eq:laplacechi}
\end{equation}
where
\begin{align}
\nonumber\chi(\lambda,s,\xi) &=\int_0^\infty d\eta e^{-\eta-\lambda\eta-s h(\eta)-\xi\theta(\eta_c-\eta)}\\
&=\frac{e^{-s}}{\lambda+1}\left[e^{-\xi}(1-e^{-\eta_c (1+\lambda)})+e^{-\eta_c (1+\lambda)}-\frac{bs}{\lambda+1+bs}e^{-\eta_c (1+\lambda)}\right]\ .
\end{align}
As for the final position, the $k$-th moment of $n_t$ after $n$ steps, conditioned on the total budget $C$, is given by
\begin{equation}
    \langle n_t^k\rangle_{C,n}=\frac{\int dn_t dX n_t^k P(n_t,X,C|n)}{\int dn_t dX  P(n_t,X,C|n)}\ ,
\end{equation}
where the denominator is simply $P(C|n)$, the marginal pdf of the total cost alone after $n$ jumps, which we computed in Eq. \eqref{PCNeq}.

\begin{figure}
    \centering
    \fbox{\includegraphics[scale = 0.5]{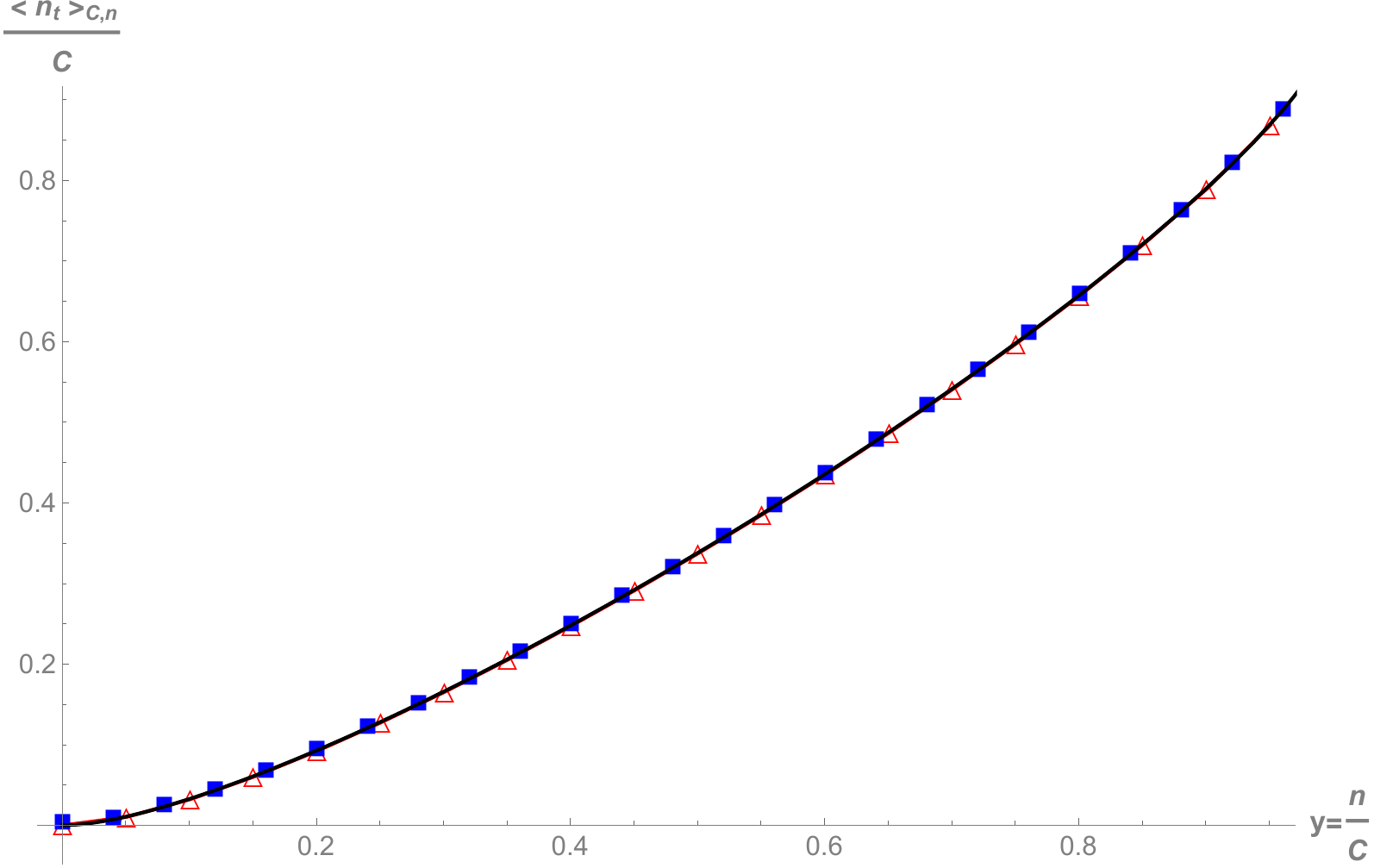}}
    \caption{Scaling behavior of the number of time-like steps $\langle n_t\rangle_{C,n}$ after $n$ steps conditioned on the total cost of the ride. On the vertical axis, we have the quantity $\langle n_t\rangle_{C,n}/C$ from Eq.~\eqref{eq:nt}. On the horizontal axis, we put $y=n/C$. We use $b=1$, $\eta_c=2$, and $C=250$ (red triangles) or $C=100$ (blue squares). We can see that the finite $n,C$ curves nicely collapse onto the scaling function $L(y)$ (solid black line) from Eq.~\eqref{eq:L}. The monotonically increasing nature of the curve here does not depend on whether $b\eta_c>1$ or $b\eta_c<1$.  }
    \label{fig:ntav}
\end{figure}

For the numerator, it is again convenient to take the Laplace transform w.r.t. the total cost, and observe from Eq. \eqref{eq:laplacechi} that
\begin{equation}
  \int dC~e^{-s C} \left[\int dn_t dX n_t^k P(n_t,X,C|n)\right]=(-1)^k\frac{\partial^k}{\partial\xi^k}\left[\chi(\lambda,s,\xi)\right]^n\Big|_{\lambda,\xi=0}\ .
\end{equation}
Limiting ourselves to the first moment ($k=1$), we obtain for the Laplace transform of the numerator
\begin{align}
    \int dC~e^{-s C} \left[\int dn_t dX n_t P(n_t,X,C|n)\right]=n e^{-ns}(1-\cosh(\eta_c)+\sinh(\eta_c))\left[1-\frac{bs e^{-\eta_c}}{1+bs}\right]^{n-1}\ .
\end{align}
Using now
\begin{equation}
 \left[1-\frac{bs e^{-\eta_c}}{1+bs}\right]^{n-1}=(1-e^{-\eta_c})^{n-1}\left[1+\frac{A}{1+bs}\right]^{n-1}\ ,
\end{equation}
where $A$ is defined in Eq.~\eqref{eq:A}, and expanding using the binomial theorem, we get
\begin{align}
 \nonumber    \int dC~e^{-s C} \left[\int dn_t dX n_t P(n_t,X,C|n)\right] &= n (1-\cosh(\eta_c)+\sinh(\eta_c))(1-e^{-\eta_c})^{n-1}\times\\
    &\times \left[e^{-ns}+\sum_{k=1}^{n-1}{n-1\choose k}A^k\frac{e^{-ns}}{(1+bs)^k}\right]\ .
\end{align}
Using the inverse Laplace transform in Eq. \eqref{eq:invLaplH}, we get for the numerator
\begin{align}
\nonumber &\int dn_t dX n_t P(n_t,X,C|n) =  n (1-\cosh(\eta_c)+\sinh(\eta_c))(1-e^{-\eta_c})^{n-1}\times\\
\nonumber  &\times\left[\delta(C-n)+\frac{e^{-(C-n)/b}}{b}\theta(C-n)\sum_{k=1}^{n-1}{n-1\choose k}\frac{A^k\left(\frac{C-n}{b}\right)^{k-1}}{\Gamma(k)}\right]\\
\nonumber &=n (1-\cosh(\eta_c)+\sinh(\eta_c))(1-e^{-\eta_c})^{n-1}\times\\
 &\times
 \left[\delta(C-n)+\frac{e^{-(C-n)/b}}{b}\theta(C-n)A(n-1)~_1 F_1(2-n,2;-A(C-n)/b)\right]\ ,
\end{align}
where we used the identity Eq. \eqref{eq:identity1kummer} in the last step. Putting everything together, we get
\begin{equation}
    \langle n_t\rangle_{C,n}=n\delta(C-n)+\theta(C-n)(n-1)\frac{_1 F_1\left(2-n,2;-\frac{A}{b}(C-n)\right)}{_1 F_1\left(1-n,2;-\frac{A}{b}(C-n)\right)}\ .
    \label{eq:nt}
\end{equation}
The delta contribution is easy to understand: if the total budget $C$ is exactly equal to $n$, then a trajectory made up of $n$ time-like steps meets all the constraints. Interestingly, the continuous part $(C>n)$ admits a scaling form for $n,C\to\infty$ such that $y=n/C$ is fixed, namely
\begin{equation}
    \langle n_t\rangle_{C,n} \sim C L\left(n/C\right)
\end{equation}
where the scaling function
\begin{align}
    L(y) &= y  \left[1+u(y)/2+ (u(y)/2) \sqrt{1+4/u(y)}\right]^{-1}\\
    u(y) &=\frac{A}{b}\left(\frac{1}{y}-1\right) \label{eq:L}
\end{align}
for $0<y<1$ simply follows from the asymptotics in Eq. \eqref{defPSI} for $k=m=2$. We recall that the costant $A$ is defined in Eq.~\eqref{eq:A}. The scaling function $L(y)$ has asymptotic behaviors
\begin{equation}
    L(y)\approx \begin{cases}
        b(e^{\eta_c}-1)y^2\quad& \text{ for }y\to0^+\,,\\
        1-\sqrt{\frac{A(1-y)}{b}}\quad& \text{ for }y\to1^-\,.
    \end{cases}
\end{equation}
A plot of the scaling function is provided in Fig.~\ref{fig:ntav}. The number of time-like steps increases monotonously with $n$ at fixed $C$. Indeed, for all values of $b$ and $\eta_c$, the scaling function $L(y)$ grows from $L(y=0)=0$ to $L(y=1)=1$. This observation agrees with the intuition that increasing the total number of steps at fixed cost leads to more ``timelike'' steps.

\section{Numerical simulations}
\label{sec:simul}
In this section, we describe the constrained Monte Carlo Markov Chain (MCMC) algorithm used to investigate the statistics of the total displacement $X$ with fixed number of steps $n$ and fixed total cost $C$. We adapt the technique used in Refs.~\cite{nadal10,nadal11,gradenigo19,mori21} to our problem. For a given initial cost $C\geq n$, we consider an $n$-dimensional vector $\vec{\eta}=\{\eta_1\,,\ldots\,,\eta_n\}$ where $\eta_i$ is the displacement at the $i$-th step. We initialize the vector choosing $\eta_1=\eta_2=\ldots=\eta_n=\eta_c+(C/n-1)/b$, such that $\sum_i h(\eta_i)=C$. We then implement a MCMC algorithm accepting only moves that do not violate the constraint $C(1-\epsilon)\leq C_{\rm new}\leq C(1+\epsilon)$, where $C_{\rm new}$ is the cost after the proposed move and we set $\epsilon=0.01$. In other words, at each iteration, 
\begin{enumerate}
    \item We choose a random integer $1\leq i\leq n$.
    \item Propose a move $\eta_i\to\eta_i^{\rm new}=\eta_i+z$, where $z$ is a uniform random variable in $[-\delta,\delta]$. We choose $\delta$ so that the acceptance rate is close to $1/2$.
    \item Evaluate the new cost $C_{\rm new}$. If $C_{\rm new}$ does not satisfy $C(1-\epsilon)\leq C_{\rm new}\leq C(1+\epsilon)$, reject the move.
    \item Otherwise, accept the move with probability $\alpha=\min[1,e^{-\eta_i^{\rm new}+\eta_i}]$. This step makes sure that the variables $\eta$ are drawn from the correct exponential distribution.
\end{enumerate}
We let the system thermalize for $10^4 n$ steps and then we sample the position $X=\sum_i\eta_i$ every $10^3n$ steps to avoid sample correlations. The results of numerical simulations are shown in the insets of Figs.~\ref{fig:CondAv} and \ref{fig:CondVar} and are in good agreement with our theoretical predictions.

\section{Conclusions}\label{sec:conclusions}

In this paper, we have computed the exact statistics of the distance covered by a one-dimensional random walk subject to a non-linear cost function: slow/short jumps of size $<\eta_c$ incur a flat fee ($=1$ unit), while long/fast jumps of size $>\eta_c$ are charged an amount proportional to the size of the jump, according to the cost function $h(\eta)$ defined in Eq.~\eqref{eq:h_eta}. The Nonlinear-Cost Random Walk with exponential jump distribution was introduced in our recent letter \cite{letterPRL}, where we focused on random walks with a constraint in the total distance $X$ and/or the total number of steps $n$. Here, we instead studied random walks that are constrained to be realized with a fixed budget $C$. We found that the average and variance of the total distance covered exhibit a rich non-monotonic behavior as a function of the scaling variable $y=n/C<1$. We also computed the statistics of the number of long/short jumps making up a trajectory of size $n$. All the analytical results have been corroborated with careful numerical simulations, obtained via a constrained Monte Carlo method that implemented the fixed-budget constraint.

There are several possible extensions of this work. First of all, it would be interesting to generalize our model to consider other nonlinear cost functions and distributions of the steps of the random walk. For instance, it would be relevant to extend our framework to Lévy walks, where condensation transitions where a single step dominates the whole trajectory can be observed \cite{majumdar2005}. Furthermore, while our current paper exclusively addresses random walks characterized by positive steps, it would be worth to explore scenarios where both positive and negative step increments are permitted. This would allow to investigate the statistics of extremes and records \cite{EVSreview}.

\subsection*{Acknowledgments}

This work was supported by a Leverhulme Trust International Professorship Grant (No. LIP-2020-014).
\appendix
\section{Variance calculation}\label{appvarA}
It is convenient to first rewrite
\begin{equation}
    g(\lambda,s)=e^{-s}[f_1(\lambda+1)+f_2(\lambda+1)+f_3(\lambda+1)]
\end{equation}
where
\begin{align}
    f_1(x) &=\frac{1}{x}\\
    f_2(x) &=-\frac{e^{-\eta_c x}}{x}\\
    f_3(x) &= \frac{e^{-\eta_c x}}{x+bs}\ ,
\end{align}
from which we can compute derivatives quite easily
\begin{align}
    f_1^{(1)}(x) &=-\frac{1}{x^2}\\
    f_2^{(1)}(x) &=\frac{e^{-\eta_c x}}{x}\left(\frac{1}{x}+\eta_c\right)\\
    f_3^{(1)}(x) &= -\frac{e^{-\eta_c x}}{x+bs}\left(\frac{1}{x+bs}+\eta_c\right)\ ,
\end{align}
and
\begin{align}
    f_1^{(2)}(x) &=\frac{2}{x^3}\\
    f_2^{(2)}(x) &=-\frac{e^{-\eta_c x}}{x}\left(\frac{2}{x^2}+\frac{2\eta_c}{x}+\eta_c^2\right)\\
    f_3^{(2)}(x) &= \frac{e^{-\eta_c x}}{x+bs}\left(\frac{2}{(x+bs)^2}+\frac{2\eta_c}{x+bs}+\eta_c^2\right)\ .
\end{align}
We can now evaluate the ratios of $g_k(s)$ appearing in \eqref{eq:X2avg} as
\begin{align}
    \frac{g_2(s)}{g_0(s)} &=\frac{f_1^{(2)}(1)+f_2^{(2)}(1)+f_3^{(2)}(1)}{f_1(1)+f_2(1)+f_3(1)}=\frac{2-e^{-\eta_c}(2+2\eta_c+\eta_c^2)+\frac{e^{-\eta_c}}{r}(2/r^2+2\eta_c/r+\eta_c^2)}{1-e^{-\eta_c}+e^{-\eta_c}/r}\\
    \frac{g_1(s)}{g_0(s)} &=\frac{f_1^{(1)}(1)+f_2^{(1)}(1)+f_3^{(1)}(1)}{f_1(1)+f_2(1)+f_3(1)}=\frac{-1+e^{-\eta_c}(1+\eta_c)-\frac{e^{-\eta_c}}{r}(1/r+\eta_c)}{1-e^{-\eta_c}+e^{-\eta_c}/r}
\end{align}
Multiplying up and down by $r e^{\eta_c}$, we get
\begin{align}
      \frac{g_2(s)}{g_0(s)} &=\frac{2e^{\eta_c}r-r(2+2\eta_c+\eta_c^2)+(2/r^2+2\eta_c/r+\eta_c^2)}{1-r+r e^{\eta_c}}\\
      \frac{g_1(s)}{g_0(s)} &=\frac{-r e^{\eta_c}+r(1+\eta_c)-1/r-\eta_c}{1-r+r e^{\eta_c}}\ .
\end{align}

\section{Calculation of $T_1(C)$ in Eq. \eqref{eq:T1Cmain}}\label{app:T1C}

We denote by
\begin{equation}
    T_1(C)=\langle X\rangle = \int_0^\infty X P(X,C|n)dX\label{eq:T1Capp}
\end{equation}
the unconstrained average of the final position of the walker after $n$ steps.

Let us start from the Laplace transform \eqref{momentLaplace} for $k=1$
\begin{equation}
    \int_0^\infty T_1(C)e^{-sC} dC=-n [g(0,s)]^{n-1}g_1(s)\ ,
\end{equation}

where $g_k(s)$ are defined in \eqref{gk}. Computing $-g_1(s)$ (see Appendix \ref{appvarA}), we eventually get
\begin{equation}
   \int_0^\infty T_1(C)e^{-sC} dC=\varphi(1+bs)\ ,\label{LaplaceT1}
\end{equation}
where 
\begin{equation}
    \varphi(r)=n e^{-n(r-1)/b}\left[1-e^{-\eta_c}+\frac{e^{-\eta_c}}{r}\right]^{n-1} \left[1-e^{-\eta_c}(1+\eta_c)+\frac{e^{-\eta_c}}{r}(1/r+\eta_c)\right]\ .
\end{equation}
Setting $1+bs=r$ in the l.h.s. of \eqref{LaplaceT1}, and setting $\hat C=C/b$, we get
\begin{equation}
    \int_0^\infty \hat T_1(\hat C)e^{-r \hat C}d\hat C=\varphi(r)\ ,\label{LaplaceT1hat}
\end{equation}
where 
\begin{equation}
    \hat T_1(\hat C)=b T_1(b \hat C)e^{\hat C}\ .\label{defhatT1}
\end{equation}
Therefore, it is convenient to inverse-Laplace transform \eqref{LaplaceT1hat} w.r.t. $r$, and then use the relation \eqref{defhatT1} to reconstruct $T_1(C)$.

In order to inverse-Laplace transform \eqref{LaplaceT1hat} w.r.t. $r$, we first rewrite
$\varphi(r)$ as
\begin{align}
 \nonumber  \varphi(r) &=ne^{n/b}(1-e^{-\eta_c})^{n-1}\left[1-e^{-\eta_c}(1+\eta_c)\right]e^{-(n/b)r}
    \left[1+\frac{A}{r}\right]^{n-1} \left[1+\frac{y}{r}+\frac{z}{r^2}\right]\\
    &=C(n,b,\eta_c)e^{-(n/b)r}\left[1+\frac{y}{r}+\frac{z}{r^2}\right]\left[1+\sum_{k=1}^{n-1}{n-1\choose k}\frac{A^k}{r^k}\right]\ ,
\end{align}
where
\begin{align}
    A &=\frac{e^{-\eta_c}}{1-e^{-\eta_c}}\label{eq:x}\\
    y &=\frac{\eta_c e^{-\eta_c}}{1-e^{-\eta_c}(1+\eta_c)}\label{eq:y}\\
    z &=\frac{e^{-\eta_c}}{1-e^{-\eta_c}(1+\eta_c)}\label{eq:z}\\
    C(n,b,\eta_c) &=ne^{n/b}(1-e^{-\eta_c})^{n-1}\left[1-e^{-\eta_c}(1+\eta_c)\right]\ .
\end{align}
Using now the elementary inverse Laplace transform for $k\geq 1$
\begin{equation}
    H_{a,k}(\hat C)=\mathcal{L}^{-1}_r\left[\frac{e^{-a r}}{r^k}\right](\hat C)=\frac{\theta (\hat C-a) (\hat C-a)^{k-1}}{\Gamma (k)}\ ,\label{eq:invLaplH}
\end{equation}

we have that
\begin{align}
\nonumber \hat T_1(\hat C) &= C(n,b,\eta_c)\left[\delta\left(\hat C-\frac{n}{b}\right)+\sum_{k=1}^{n-1}{n-1\choose k}A^k H_{n/b,k}(\hat C)+y\sum_{k=1}^{n-1}{n-1\choose k}A^k H_{n/b,k+1}(\hat C)\right.\\
 &\left. +y H_{n/b,1}(\hat C) +z\sum_{k=1}^{n-1}{n-1\choose k}A^k H_{n/b,k+2}(\hat C)+z H_{n/b,2}(\hat C) \right]\ .
\end{align}
The finite sums can be further computed in closed form using
\begin{align}
    \sum_{k=1}^{n-1}{n-1\choose k}\frac{A^k}{\Gamma(k)} &=A (n-1) \, _1F_1(2-n;2;-A)\label{eq:identity1kummer}\\
    \sum_{k=1}^{n-1}{n-1\choose k}\frac{A^k}{\Gamma(k+1)} &=\, _1F_1(1-n;1;-A)-1\\
     \sum_{k=1}^{n-1}{n-1\choose k}\frac{A^k}{\Gamma(k+2)}&=\, _1F_1(1-n;2;-A)-1\ ,
\end{align}
leading to
\begin{align}
\nonumber \hat T_1(\hat C) &= C(n,b,\eta_c)\left[\delta\left(\hat C-\frac{n}{b}\right)+\theta(\hat C-n/b)\left[(n-1)A~ _1F_1(2-n;2;-A(\hat C-n/b))\right. \right.\\
&\left.\left.+y~ _1F_1(1-n;1;-A(\hat C-n/b))+z(\hat C-a)~ _1F_1(1-n;2;-A(\hat C-n/b)) \right]\right]\ .
\end{align}

Using the relation \eqref{defhatT1}, we eventually obtain Eq. \eqref{eqT1C} of the main text.

\section{Calculation of $T_2(C)$ in Eq. \eqref{eq:T2Cmain}}\label{app:T2C}

We denote by
\begin{equation}
    T_2(C)=\langle X^2\rangle = \int_0^\infty X^2 P(X,C|n)dX\label{eq:T2app}
\end{equation}
the unconstrained second moment of the final position of the walker after $n$ steps.

Let us start again from the Laplace transform \eqref{momentLaplace} for $k=2$
\begin{equation}
    \int_0^\infty T_2(C)e^{-sC} dC=n(n-1)g_0(s)^{n-2}g_1^2(s)+n g_0(s)^{n-1}g_2(s)\ ,
\end{equation}

where $g_k(s)$ are defined in \eqref{gk}. Computing $g_0(s),g_1(s),g_2(s)$ (see appendix \ref{appvarA}), we eventually get
\begin{equation}
   \int_0^\infty T_2(C)e^{-sC} dC=\Psi(1+bs)\ ,
\end{equation}
where 
\begin{align}
\nonumber    \Psi(r) &=n(n-1)e^{n/b}e^{-(n/b)r}(1-e^{-\eta_c})^{n-2}(e^{-\eta_c}(1+\eta_c)-1)^2\left[1+\frac{A}{r}\right]^{n-2}\left[1-\frac{y}{r}-\frac{z}{r^2}\right]^2\\
    &+ne^{n/b}e^{-(n/b)r}(1-e^{-\eta_c})^{n-1}(2-e^{-\eta_c}(2+2\eta_c+\eta_c^2))\left[1+\frac{A}{r}\right]^{n-1}\left[1+\frac{t}{r^3}+\frac{\phi}{r^2}+\frac{\xi}{r}\right]\ ,\label{eq:Psir}
\end{align}
with 
\begin{align}
A &= \frac{\exp (-\eta_c )}{1-\exp (-\eta_c )}\\
y&= \frac{\eta_c  \exp (-\eta_c )}{(\eta_c +1) \exp (-\eta_c )-1}\\
z&= \frac{\exp (-\eta_c )}{(\eta_c +1) \exp (-\eta_c )-1}\\
    t &= \frac{2e^{-\eta_c}}{2-e^{-\eta_c}(2+2\eta_c+\eta_c^2)}\\
    \phi &=\frac{2\eta_c e^{-\eta_c}}{2-e^{-\eta_c}(2+2\eta_c+\eta_c^2)}\\
     \xi &=\frac{\eta_c^2 e^{-\eta_c}}{2-e^{-\eta_c}(2+2\eta_c+\eta_c^2)}\ .
\end{align}
(Note the different definition of $A,y,z$ here w.r.t. Eqs. \eqref{eq:x}, \eqref{eq:y} and \eqref{eq:z}).

Setting $1+bs=r$ in the l.h.s. of \eqref{LaplaceT1}, and setting $\hat C=C/b$, we get
\begin{equation}
    \int_0^\infty \hat T_2(\hat C)e^{-r \hat C}d\hat C=\Psi(r)\ ,\label{LaplaceT2hat}
\end{equation}
where 
\begin{equation}
    \hat T_2(\hat C)=b T_2(b \hat C)e^{\hat C}\ .\label{defhatT2}
\end{equation}
Therefore, it is convenient to inverse-Laplace transform Eq. \eqref{eq:Psir} w.r.t. $r$, and then use the relation \eqref{defhatT2} to reconstruct $T_2(C)$.

We now define for convenience (for $k>0$)
\begin{equation}
    H_{a,m,k}(\hat C)=\mathcal{L}^{-1}\left[e^{-ar}\left(1+\frac{A}{r}\right)^m\frac{1}{r^k}\right](\hat C)=\frac{\theta(\hat C-a)(\hat C-a)^{k-1}~ _1F_1(-m,k;-A(\hat C-a))}{\Gamma(k)}
\end{equation}
and for $k=0$
\begin{equation}
    H_{a,m,0}(\hat C)=\mathcal{L}^{-1}\left[e^{-ar}\left(1+\frac{A}{r}\right)^m\right](\hat C)=\delta(\hat C-a)+mA~_1F_1(1-m,2;-A(\hat C-a))\theta(\hat C-a)\ .
\end{equation}

In terms of these auxiliary functions, the inverse Laplace transform reads
\begin{align}
 \nonumber   \hat T_2(\hat C) &=M(n,b,\eta_c)\left\{H_{n/b,n-2,0}(\hat C)-2yH_{n/b,n-2,1}(\hat C)+(y^2-2z)H_{n/b,n-2,2}(\hat C)+2yz H_{n/b,n-2,3}(\hat C)\right.\\
 \nonumber    &\left.+z^2 H_{n/b,n-2,4}(\hat C)\right\}+\Theta(n,b,\eta_c)\left\{H_{n/b,n-1,0}(\hat C)+tH_{n/b,n-1,3}(\hat C)+\phi H_{n/b,n-1,2}(\hat C)\right.\\
    &\left.+ \xi H_{n/b,n-1,1}(\hat C)\right\}\ ,\label{eq:t2hat}
\end{align}
where
\begin{align}
M(n,b,\eta_c) &=  n(n-1)e^{n/b}(1-e^{-\eta_c})^{n-2}(e^{-\eta_c}(1+\eta_c)-1)^2\\
 \Theta(n,b,\eta_c) &=ne^{n/b}(1-e^{-\eta_c})^{n-1}(2-e^{-\eta_c}(2+2\eta_c+\eta_c^2))\ .
\end{align}
Finally, we find from \eqref{defhatT2}
\begin{equation}
T_2(C)=\frac{1}{b}e^{-C/b}\hat T_2(C/b)\,,\label{eq:T2C}
\end{equation}
where $\hat T_2(\hat C)$ is given in Eq.~\eqref{eq:t2hat}.

\section{Asymptotics of ratio of hypergeometric functions}\label{app:ratiohyp}

We compute here the asymptotic behavior for large $n$ of the ratio of Kummer hypergeometric functions appearing in the definition of $R(k,n,m,u)$ (see Eq. \eqref{eq:ratioKummer}), where both the first and last argument depend on $n$:

\begin{equation}
    \Psi_{m,k}(u)=\lim_{n\to \infty} \left[n^{m-2} \frac{ _1 F_1(k-n, m, -n u)}{_1 F_1(1-n,2,-nu)}\right] = \Gamma(m) u^{1-m/2} \left[1+u/2+ (u/2) \sqrt{1+4/u}\right]^{m/2-k}\ .\label{defPSI}
\end{equation}

We need the identities
\begin{equation}
     _0 F_1(m,z)= \Gamma(m) I_{m-1}( 2 \sqrt{z})/z^{(m-1)/2}\ ,
\end{equation}
where $I_m(z)$ is a Bessel function, and (for $\mathrm{Re}(a)>0$)

\begin{equation}
    _1 F_1(a,b;z)=\frac{1}{\Gamma(a)}\int_0^\infty dt~e^{-t}t^{a-1}~ _0 F_1(b,zt)\ ,
\end{equation}
as well as
\begin{equation}
    _1 F_1(a,b;-z)=e^{-z}~_1 F_1(b-a,b;z)\ .
\end{equation}

Therefore
\begin{align}
\nonumber _1 F_1(k-n, m, -n u) &=e^{-nu}~_1 F_1(m-k+n, m, n u)=\\
& e^{-nu}\frac{\Gamma(m)}{\Gamma(m-k+n)n^{(m-1)/2}} \int_0^\infty dt~e^{-t}t^{m-k+n-1}  \frac{I_{m-1}( 2 \sqrt{ntu})}{u^{(m-1)/2} t^{(m-1)/2}}\ .
\end{align}
Using the asymptotic behavior of the Bessel function for large argument
\begin{equation}
    I_\alpha(z)\sim \frac{e^z}{\sqrt{2\pi z}} 
\end{equation}
combined with the change of variable $t=n\tau$, we get
\begin{equation}
 _1 F_1(k-n, m, -n u) \sim \frac{e^{-nu}}{2\sqrt{\pi}~u^{\frac{m}{2}-\frac{1}{4}}}\frac{\Gamma(m)}{\Gamma(m-k+n)}n^{n-k+1/2}\int_0^\infty d\tau~ \tau^{m/2-k-3/4}e^{-n W(\tau,u)}\ , \label{SPKummer}
\end{equation}
with
\begin{equation}
    W(\tau,u)=\tau-\ln\tau-2\sqrt{\tau u}\ .
\end{equation}
The integral in \eqref{SPKummer} can be evaluated using a saddle point approximation for large $n$. The only critical value inside the integration interval is at
\begin{equation}
    \frac{dW}{d\tau}\Big|_{\tau^\star}=0\Rightarrow \tau^\star -\sqrt{u}\sqrt{\tau^\star}-1=0\Rightarrow \tau^\star(u)=\frac{u}{2}+1+\frac{u}{2}\sqrt{1+\frac{4}{u}}\ .
\end{equation}
Therefore
\begin{equation}
\int_0^\infty d\tau~ \tau^{m/2-k-3/4}e^{-n W(\tau,u)}\sim \sqrt{\frac{2\pi}{n |W''(\tau^\star(u),u)|}} [\tau^\star (u)]^{m/2-k-3/4}e^{-n W(\tau^\star(u),u)}\ ,   
\end{equation}
from which it follows that the ratio of Kummer hypergeometric functions in \eqref{defPSI} (where for the denominator we simply set $k=1$ and $m=2$) simplifies dramatically for large $n$ as
\begin{equation}
    \frac{ _1 F_1(k-n, m, -n u)}{_1 F_1(1-n,2,-nu)}\sim \Gamma(m) u^{1-m/2}n^{1-k}\frac{\Gamma(1+n)}{\Gamma(m-k+n)\Gamma(2)}[\tau^\star(u)]^{m/2-k}\ .
\end{equation}
The result in \eqref{defPSI} then follows by noting that $\Gamma(2)=1$ and $\Gamma(1+n)/\Gamma(a+n)\sim n^{1-a}$ for large $n$. Starting from Eq. \eqref{eq:avXconstrainedfinal} and replacing every occurrence of $R$ with its corresponding asymptotic behavior in \eqref{defPSI} yields, after simplification, the same scaling relation found in \eqref{finalscaling} with the inverse-Laplace method.

\end{document}